\documentclass[10pt]{article}
\usepackage{arxiv}
\usepackage{amsmath}
\usepackage{graphicx}
\usepackage{geometry}
\usepackage{authblk}
\usepackage{dashrule}
\usepackage[table]{xcolor}

\usepackage{tikz}
\usetikzlibrary{shapes,arrows}
\usetikzlibrary{trees}
\usepackage{multirow}

\usepackage[justification=centering]{caption}

\usepackage[
  style=numeric-comp,
  datamodel=software, % extend the datamodel with entries for software
  abbreviate=false,
  natbib=true,
  sorting=none,
  backend=bibtex,
  bibencoding=utf8,
  url=true,
  doi=true,
  defernumbers,
  maxcitenames=10,
  defernumbers=false,
  maxbibnames=100]{biblatex}

\newcolumntype{P}[1]{>{\centering\arraybackslash}p{#1}}

\colorlet{myColor}{blue!0!black}
\usepackage[colorlinks=true,linkcolor=myColor,citecolor=myColor,urlcolor=myColor]{hyperref}

\usepackage{mathptmx} % Use Times New Roman font
\usepackage{titlesec}

\titleformat{\section}{\centering\bfseries\fontsize{10}{12}\selectfont}{\thesection}{1em}{}
\titleformat{\subsection}{\bfseries\fontsize{10}{12}\selectfont}{\thesubsection}{1em}{}

\usepackage{fancyhdr}
\fancypagestyle{titlepage}{
    \fancyhf{} % Clear all header and footer fields
    \fancyhead[L]{ \textbf{International Journal Of Engineering Research And Development} \\
    e- ISSN: 2278-067X, p-ISSN: 2278-800X, \underline{www.ijerd.com} \\
    Volume 20, Issue 9 (September, 2024), PP.08-23}
		\fancyfoot[C]{
    \hrule
    \vspace{2pt} % Adjusts the space between the line and the footer text
    \thepage
}
}

\title{{Enhancing 5G-NR mmWave : Phase Noise Models Evaluation with MMSE  for CPE Compensation}}

\author[1,*]{Desire~Guel}
\author[1]{Flavien~Herve~Somda}
\author[2]{Boureima~Zerbo}
\author[3]{Oumarou~Sie}
\affil[1]{University of Joseph KI-ZERBO (U-JKZ), Burkina Faso}
\affil[2]{University of Thomas SANKARA (UTS), Burkina Faso}
\affil[3]{Aube Nouvelle University (U-AUBEN), Burkina Faso}
\affil[*]{Corresponding Author: guel.desire@gmail.com}

\date{}

\begin{document}

\maketitle

\thispagestyle{titlepage} % Apply the custom style only to the first page
\pagestyle{fancy}
\fancyhf{}
\fancyhead[L]{}
\fancyhead[R]{\textsl{Enhancing 5G-NR mmWave : Phase Noise Models Evaluation with MMSE  for CPE Compensation}}
%\fancyfoot[C]{\thepage}
\fancyfoot[C]{
    \hrule
    \vspace{2pt} % Adjusts the space between the line and the footer text
    \thepage
}

\noindent \hrule
\vspace*{+10pt}
%\section*{ABSTRACT}
%\section*{\raggedright\textbf{ABSTRACT}}
\noindent \textbf{ABSTRACT - }

The rapid development of 5G New Radio (NR) and millimeter-wave (mmWave) communication systems highlights the critical importance of maintaining accurate phase synchronization to ensure reliable and efficient communication. This study focuses on evaluating phase noise models and implementing Minimum Mean Square Error (MMSE) algorithms for Common Phase Error (CPE) compensation. Through extensive simulations, we demonstrate that CPE compensation significantly enhances signal quality by reducing Error Vector Magnitude (EVM) and Bit Error Rate (BER) across various Signal-to-Noise Ratio (SNR) levels and antenna configurations. Results indicate that implementing MMSE-based CPE estimation and compensation  in 5G-NR mmWave systems reduced EVM from 7.4\% to 4.6\% for 64QAM and from 5.4\% to 4.3\% for 256QAM, while also decreasing BER from $5.5 \times 10^{-3}$ to $5.2 \times 10^{-5}$ for 64QAM, demonstrating significant improvements in signal quality and reliability across various SNR levels and antenna configurations. Our findings provide valuable insights for optimizing phase noise mitigation strategies in 5G-NR mmWave systems, contributing to the development of more robust and efficient next-generation wireless networks.

%\vspace*{-5pt}
\noindent \hdashrule[0.5ex]{1.\linewidth}{0.25pt}{1mm 0.5mm} % {line width}{dash length}{gap length}
\noindent Date of Submission: 27-08-2024 \hspace*{+190pt} Date of acceptance: 05-09-2024 \\
\noindent \hdashrule[0.5ex]{1.\linewidth}{0.25pt}{1mm 0.5mm} % {line width}{dash length}{gap length}

\section{INTRODUCTION}
\label{sec:Introduction}

Phase synchronization is a critical component in the performance of 5G New Radio (NR) Millimeter-wave (mmWave) communication systems. Accurate phase synchronization is essential for maintaining the reliability and efficiency of communication, particularly within the mmWave frequency bands, which typically range from 24 GHz to 100 GHz. These high-frequency bands enable unprecedented data rates and bandwidth which are vital for meeting the increasing demand for high-speed wireless connectivity. The evolution of 5G-NR relies heavily on mmWave technology to provide enhanced mobile broadband services, ultra-reliable low-latency communication and massive machine-type communication, addressing the capacity constraints of traditional frequency bands \cite{Naqvi2021,Dikarev2022,ShahenShah2022}.

However, the deployment of 5G-NR mmWave networks is accompanied by significant challenges, particularly in the accurate estimation and compensation of phase errors. These errors arise from various sources, including oscillator imperfections, channel effects and hardware impairments, all of which can induce Common Phase Error (CPE). CPE estimation and compensation are crucial for ensuring reliable communication in mmWave systems, as even minor phase deviations can significantly degrade system performance, leading to increased error rates and reduced signal quality \cite{Abhayawardhana2002}.

Despite ongoing research, there remains a gap in the comprehensive evaluation of phase noise models and their effectiveness in improving CPE compensation, particularly using advanced algorithms like Minimum Mean Square Error (MMSE). While previous studies have explored various phase noise models and compensation techniques, the growing complexity and stringent performance requirements of 5G-NR mmWave systems necessitate a more thorough evaluation of these models. Specifically, there is a need to assess the effectiveness of phase noise models when combined with MMSE algorithms, a challenge that has not been fully addressed in the current literature.

This study seeks to fill this gap by providing a comprehensive evaluation of phase noise models and applying MMSE algorithms for CPE estimation and compensation in 5G-NR mmWave systems. In contrast to earlier works, this research introduces a novel evaluation framework specifically designed to address the unique challenges of the mmWave spectrum. Through this framework, we conduct an in-depth analysis of commonly used phase noise models and compare the performance of  MMSE algorithm in mitigating phase errors. The findings of this study are expected to contribute valuable insights into the practical optimization of phase noise models and compensation techniques, ultimately enhancing the reliability and efficiency of 5G-NR mmWave communication systems.

The remainder of this article is structured as follows: In Section \ref{sec:LiteratureReview}, we conduct a detailed literature review to position our work within the broader context of existing research. Section \ref{sec:SystemModel} describes the system model used in our evaluation, detailing the performance assessment of various phase noise models and the implementation of state-of-the-art algorithms for CPE estimation and compensation. Section \ref{sec:Methodology} outlines the experimental setup and methodologies employed in this study. Finally, Section \ref{sec:ResultsAndDiscussion} presents the results, followed by a comprehensive discussion of the findings, highlighting the strengths and limitations of the evaluated models and algorithms and providing insights for future research directions.

\section{LITERATURE REVIEW}
\label{sec:LiteratureReview}

In this section, we  review the existing research on phase noise models, Minimum Mean Square Error (MMSE) algorithms and Common Phase Error (CPE) compensation techniques in millimeter-wave (mmWave) communication systems. The review focuses on the strengths and limitations of these studies, highlighting the need for a comprehensive evaluation framework that addresses the challenges posed by 5G NR mmWave environments. 

Table \ref{tab:relatedworks} provides a comprehensive taxonomy of the related works, summarizing key contributions, strengths and limitations of existing studies on phase noise models, MMSE algorithms and CPE compensation techniques in mmWave communication systems.

\begin{table*}[h!]
\centering
\caption{Taxonomy of Related Works}
\resizebox{0.85\textwidth}{!}{
\begin{tabular}{|p{2cm}|p{3cm}|p{3cm}|p{3cm}|P{1.5cm}|}
\hline
\textbf{Authors } & \textbf{Short Description} & \textbf{Strengths} & \textbf{Limitations} & \textbf{Reference} \\ \hline
Dikarev et al. (2022) & Phase noise compensation algorithm robust to timing errors. & Practical phase noise mitigation using PT-RS. & Specific to DFT-s-OFDM waveform. & \cite{Dikarev2022} \\ \hline
Park et al. (2023) & CPE estimation using MMSE equalizer. & Insights into CPE estimation techniques. & Focus on only MMSE algorithm & \cite{Park2023} \\ \hline
Azzahhafi et al. (2024) & 4x4 MIMO optical architecture for mmWave indoor applications. & High-capacity, reliable communication using RoF technology. & Limited to indoor 60 GHz unlicensed bands. & \cite{Azzahhafi2024} \\ \hline
Qi et al. (2018) & PT-RS design for phase noise tracking in 5G NR. & Foundational insights into PT-RS design. & Does not address MMSE algorithm. & \cite{Qi2018} \\ \hline
Samsung (2016) & Phase noise modeling in mmWave communication systems. & Realistic oscillator phase noise modeling. & No direct focus on compensation algorithms. & \cite{Samsung2016} \\ \hline
Gu et al. (2019) & CPE and ICI estimation in OFDM systems. & Accurate CPE and ICI cancellation methods. & Focus on adjacent sub-carrier interference. & \cite{Gu2019} \\ \hline
Kim et al. (2021) & ML-based phase noise compensation in mmWave systems. & Enhances system robustness using ML techniques. & High computational complexity. & \cite{Kim2021} \\ \hline
Zhang et al. (2022) & Phase noise modeling for 5G mmWave. & Detailed modeling for high-frequency bands. & Does not focus on compensation techniques. & \cite{Zhang2022} \\ \hline
\end{tabular}
}
\label{tab:relatedworks}
\end{table*}

Dikarev et al. \cite{Dikarev2022} address the performance degradation caused by phase noise in mmWave systems and propose a novel phase noise compensation algorithm and phase tracking reference signal (PT-RS) structure designed to be robust against timing errors. Their work underscores the critical role of PT-RS in mitigating the impact of phase noise, providing practical solutions for enhancing 5G NR uplink transmission performance. However, the proposed algorithm and PT-RS structure are tailored specifically to discrete-Fourier-transform spread orthogonal frequency division multiplexing (DFT-s-OFDM) waveforms, limiting the generalizability of their findings to other modulation schemes or communication scenarios. This specificity highlights the need for broader evaluations across different modulation techniques to enhance the applicability of phase noise compensation strategies.

Park et al. \cite{Park2023} introduce a scheme for CPE estimation using MMSE equalizers, emphasizing the importance of accurate CPE estimation for effective channel equalization. Although their study provides valuable insights into CPE estimation techniques, it primarily focuses on MMSE equalization rather than the broader evaluation of phase noise models. This limitation suggests that while their work is relevant, a more comprehensive approach that includes phase noise model evaluation is necessary to fully understand the effectiveness of MMSE algorithms in 5G NR mmWave systems.

Azzahhafi et al. \cite{Azzahhafi2024} present a novel 4x4 MIMO optical architecture for 5G mmWave indoor applications, utilizing Radio over Fiber (RoF) technology to support high-capacity and reliable communication. Although their work contributes to the design and deployment of mmWave communication systems, it focuses more on the physical layer architecture rather than on phase noise mitigation strategies. The emphasis on indoor applications at 60 GHz in unlicensed bands may limit the generalizability of their findings to other deployment scenarios. Additionally, the complexity of their proposed system poses challenges for practical deployment and cost-effectiveness, highlighting the need for more versatile and scalable solutions in phase noise compensation.

Qi et al. \cite{Qi2018} explore the effects of phase noise in mmWave bands and propose comprehensive solutions for phase noise tracking using PT-RS. Their study evaluates various PT-RS design aspects and addresses practical challenges in phase noise mitigation within 5G NR systems. Although their focus is on PT-RS design rather than on MMSE algorithms, their insights are foundational for understanding phase noise tracking techniques in 5G NR systems. This foundational work sets the stage for further investigation into comprehensive CPE estimation and compensation strategies, particularly those involving advanced algorithmic approaches.

Samsung \cite{Samsung2016} discusses phase noise modeling and its impact on orthogonal frequency-division multiplexing (OFDM) subcarrier spacing, emphasizing the need for realistic oscillator phase noise models in mmWave systems. Their discussion is crucial for understanding the challenges associated with phase noise in mmWave communication system design. However, the study does not directly address the application of  MMSE algorithms for CPE compensation, focusing instead on the broader implications of phase noise modeling. While Samsung's contribution provides valuable background information, there remains a need for studies that directly evaluate the integration of phase noise models with advanced compensation algorithms.

These articles \cite{Dikarev2022,Park2023,Azzahhafi2024,Qi2018,Samsung2016, Gu2019,Zhang2022} collectively contribute to the current understanding of phase tracking techniques, CPE estimation and compensation in mmWave communication systems. Despite the progress made, significant gaps remain, particularly in the comprehensive evaluation of phase noise models and the application of algorithmic approaches such as MMSE for CPE estimation and compensation. This study aims to address these gaps by proposing a systematic evaluation framework and implementing state-of-the-art MMSE algorithms for phase noise mitigation in 5G NR mmWave systems. The proposed research is designed to enhance the practical applicability of phase noise models and compensation techniques, ultimately improving the reliability and efficiency of next-generation communication networks.

\section{SYSTEM MODEL}
\label{sec:SystemModel}

This section provides a comprehensive analysis of the system model, encompassing an overview of the 5G-NR mmWave communication system, an exploration of the Phase Noise Model, an in-depth examination of  MMSE-based CPE Estimation and Correction techniques and the evaluation framework for Phase Noise models.

\subsection{An overview of 5G-NR mmWave communication system }
The 5G-NR mmWave communication system \cite{Naqvi2021,Zhang2022,Bogale2017} exploits the radio frequency spectrum with wavelengths ranging between 24 GHz and 100 GHz \cite{ShahenShah2022}. This specific frequency range is relatively underutilized. This offers a significant opportunity to enhance available bandwidth. Unlike lower frequencies which experience congestion due to existing TV and radio transmissions and 4G LTE networks, 5G-NR mmWave technology \cite{Naqvi2021,Zhang2022,Bogale2017} enables faster data transmission, albeit with a reduced range. The high-frequency mmWave bands, especially those exceeding 24 GHz, have been selected for their ability to support substantial bandwidths and high data rates, making them pivotal for expanding wireless network capacity.

\begin{figure}[htbp]
	\centering
	\includegraphics[page=1,clip,trim=0.0cm 0.0cm 0.0cm 0.0cm,width=0.95\textwidth]{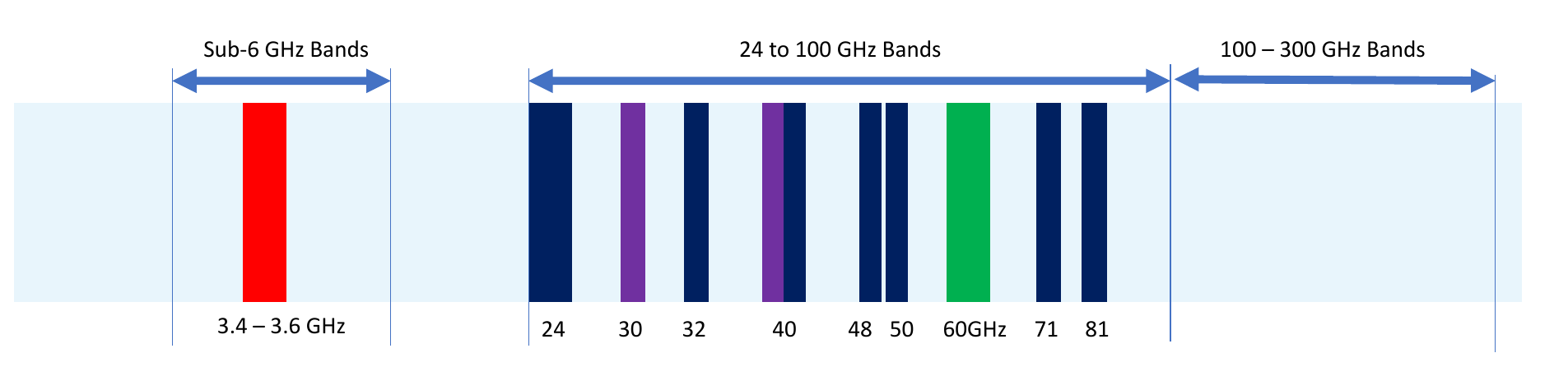}
	\caption{5G-NR mmWave spectrum \cite{ShahenShah2022}}
	\label{Fig:01}
\end{figure}

The mmWave bands between 24 GHz and 100 GHz, targeted for 5G deployment, can support bandwidths up to 2 GHz without requiring band combination, as illustrated in Figure  \ref{Fig:01}. This vast bandwidth allows 5G-NR mmWave systems to support various high-bandwidth applications, including immersive virtual reality, augmented reality, high-definition video streaming and ultra-fast file downloads.

5G-NR mmWave propagation presents unique challenges due to increased path loss, susceptibility to atmospheric absorption and sensitivity to obstacles. To overcome these challenges, 5G-NR mmWave systems employ Phase Tracking Reference Signals (PT-RS) which play a crucial role in phase synchronization between the transmitter and receiver, ensuring reliable communication. PT-RS enables the estimation and compensation of phase errors induced by various sources, including oscillator imperfections, channel effects and hardware impairments, thereby enhancing the robustness and reliability of communication links.

\subsection{Phase Noise Model}
In the realm of 5G-NR mmWave communication systems, accurate modeling of phase noise is pivotal for understanding and addressing signal degradation. Phase noise originates from various sources such as local oscillators, phase-locked loops (PLL) and hardware imperfections, posing significant challenges to reliable communication at millimeter-wave frequencies. In \cite{D5.12016}, a model considering three main noise sources—reference clock, PLL and VCO—is proposed.

We present three phase noise models: 'A' \cite{Samsung2016}, 'B' \cite{Samsung2016} and 'C' (TR 38.803 \cite{38.8032022}) which are used to capture and represent the characteristics of phase noise in the context of 5G-NR mmWave. The phase noise characteristic is expressed as follows:

\begin{equation}
S\left( f \right) = {\rm{PS}}{{\rm{D}}_{\rm{0}}}\frac{{\prod\limits_{n = 1}^N {1 + \left( {\frac{f}{{{f_{z,n}}}}} \right){\alpha _{z,n}}} }}{{\prod\limits_{m = 1}^M {1 + \left( {\frac{f}{{{f_{p,m}}}}} \right){\alpha _{p,m}}} }}
	\label{Equ:01}
\end{equation}

where ${\rm{PSD}}{_{\rm{0}}}$ is the power spectral density at zero frequency ($f = 0$) in dBc/Hz, $f_{z,n}$ are zero frequencies and $f_{p,n}$ are pole frequencies. Table \ref{Table:01} shows three parameter sets obtained from practical oscillators operating at 30 GHz, 60 GHz and 29.55 GHz, respectively \cite{Samsung2016, Qi2018}.

\begin{table}[htbp]
\centering
\caption{Parameter sets for phase noise models \cite{Samsung2016, Qi2018}.}
\label{Table:01}
\resizebox{.65\textwidth}{!}{
\begin{tabular}{|l|c|c|c|} \hline
                  & \textbf{Parameter Set-A} & \textbf{Parameter Set-B} & \textbf{Parameter Set-C} \\ \hline
Carrier frequency & 30 [GHz]                 & 60 [GHz]                 & 29.55 [GHz]             \\ 
PSD0 (dBc/Hz)     & -79.4                    & -70                      & 32                      \\ 
Fp (MHz)          & [0.1, 0.2, 8]            & [0.005, 0.4, 0.6]        & [1, 1.6, 30]            \\ 
Fz (MHz)          & [1.8, 2.2, 40]           & [0.02, 6, 10]            & [0.003, 0.55, 280]      \\ 
$\alpha_z$        & [2, 2, 2]                & [2, 2, 2]                & [2.37, 2.7, 2.53]       \\ 
$\alpha_p$        & [2, 2, 2]                & [2, 2, 2]                & [3.3, 3.3, 1]           \\ \hline
\end{tabular}
}
\end{table}

\begin{figure}[htbp]
	\centering
	\includegraphics[page=1,clip,trim=0.0cm 0.0cm 0.0cm 0.0cm,width=0.75\textwidth]{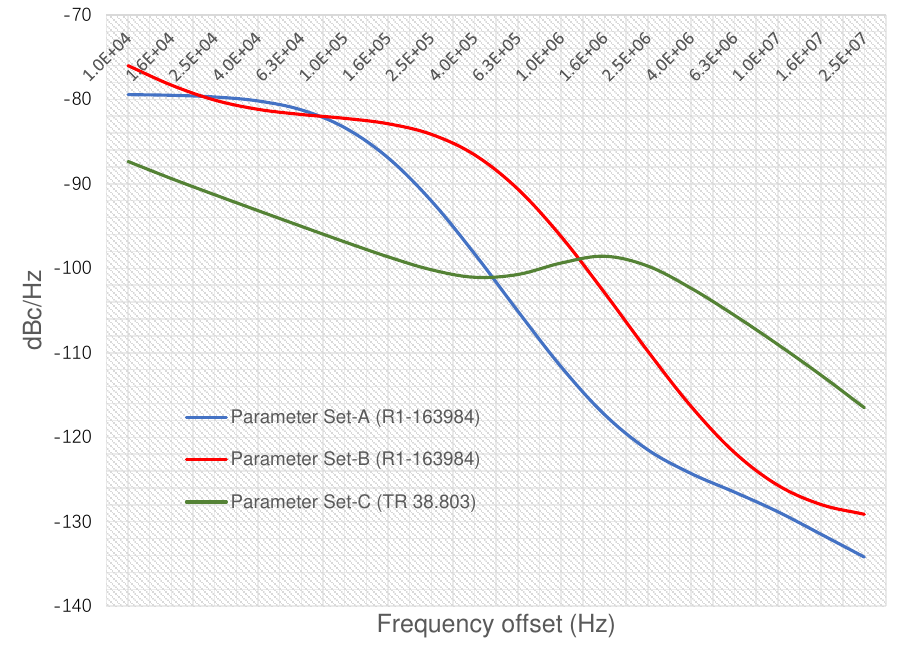}
	\caption{Phase Noise Power Spectral Density \cite{Samsung2016, Qi2018}.}
	\label{Fig:02}
\end{figure}

Figure \ref{Fig:02} shows the Phase Noise Power Spectral Density (dBc/Hz) based on the frequency (Hz) of the three phase noise models: 'A' \cite{Samsung2016}, 'B' \cite{Samsung2016} and 'C' (TR 38.803) \cite{38.8032022}.

\subsection{MMSE based CPE estimation and correction}
Accurate estimation and compensation of Common Phase Error (CPE) in 5G-NR mmWave systems require sophisticated algorithms capable of mitigating the adverse effects of phase noise. In this section, we focus on the implementation of  state-of-the-art algorithm, namely  Minimum Mean Square Error (MMSE) \cite{Apanasenko2018,Sofien2023,Garg2017} for CPE estimation and compensation.

Consider a 5G-NR communication system operating in the mmWave frequency range. Let $\mathbf{r}$ be the received signal corrupted by CPE and $\mathbf{s}$ be the transmitted signal. The received signal can be modeled as:

\begin{equation}
\mathbf{r} = \mathbf{H} \times \mathbf{s} + \mathbf{n}
\label{Equ:02}
\end{equation}

where $\mathbf{H} = \left[ e^{j\phi_{l,n}} \right]_{(l,n)}$ represents the CPE matrix and $\mathbf{n}$ is the additive white Gaussian noise vector. The MMSE equalization for the CPE matrix $\mathbf{W}_{\text{MMSE}}$ is computed as follows:

\begin{equation}
\mathbf{W}_{\text{MMSE}} = \left( \mathbf{H}^{H} \mathbf{H} + \frac{\sigma^2}{\gamma} \mathbf{I} \right)^{-1} \mathbf{H}^{H}
\label{Equ:03}
\end{equation}

where $\sigma^2$ is the noise variance, $\gamma$ is the signal-to-noise ratio (SNR) and $\mathbf{I}$ is the identity matrix.

\subsection{Phase Noise models evaluation framework}

Figure \ref{Fig:03} shows the processing chain implemented for the phase noise models evaluation in 5G-NR mmWave systems. Accurate phase noise modeling enables the identification and mitigation of phase errors, ensuring the robustness of communication links and optimizing the overall precision of 5G-NR mmWave systems. 

\tikzstyle{block} = [rectangle, draw=blue!70, fill=blue!10, 
    text width=10em, text centered, rounded corners, minimum height=3em]
\tikzstyle{block1} = [rectangle, draw=red!70, fill=red!10, 
    text width=10em, text centered, rounded corners, minimum height=3em]
\tikzstyle{line} = [draw=blue!70, -latex']
\begin{figure}[htbp]
	\centering
\begin{tikzpicture}[node distance = 1.5cm, auto]
    % Place nodes
    \node [block] (encoding) {NR-PDSCH Encoding};
    \node [block, below of=encoding] (modulation) {CP-OFDM Modulation};
    \node [block, below of=modulation] (channel) {CDL Channel + Noise};
    \node [block1, below of=channel] (noise) {Phase Noise Modeling};
    \node [block, below of=noise] (receiver) {NR-PDSCH Receiver};
    % Draw edges
    \path [line] (encoding) -- (modulation);
    \path [line] (modulation) -- (channel);
    \path [line] (channel) -- (noise);
    \path [line] (noise) -- (receiver);
\end{tikzpicture}
	\caption{NR-PDSCH Transmitter and Receiver Framework including Phase Noise.}
	\label{Fig:03}
\end{figure}
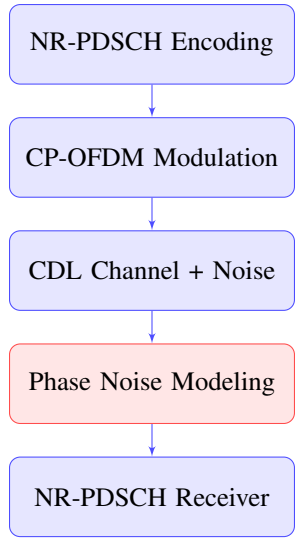

The different steps for NR-PDSCH transmission and reception (refer to Figure  \ref{Fig:03}) are described as follows:
\begin{description}
	\item [(a)] \textsl{NR-PDSCH Encoding:} In this step, the data intended for transmission over the downlink channel is encoded using error correction coding schemes such as LDPC (Low-Density Parity-Check) or Polar codes. Channel coding aims to introduce redundancy into the transmitted data to enhance error resilience and facilitate reliable communication.
	\item [(b)] \textsl{CP-OFDM Modulation:} After encoding, the modulated symbols are mapped onto resource elements (REs) within the allocated resource blocks. CP-OFDM (Cyclic Prefix Orthogonal Frequency Division Multiplexing) is employed for modulation. The modulated symbols are then transformed into the time domain using an inverse fast Fourier transform (IFFT), followed by the addition of a cyclic prefix to mitigate intersymbol interference (ISI) caused by multipath propagation.
	\item [(c)] \textsl{Phase Noise Modeling:} Phase noise is inserted to accurately represent the impact of local oscillator imperfections on the transmitted signal.
	\item [(d)] \textsl{Channel (CDL/TDL):} The modulated signal traverses through the wireless channel which can be modeled using either a Clustered Delay Line (CDL) \cite{Barb2019} or a Tap Delay Line (TDL) \cite{Barb2019} model.
	\item [(e)] \textsl{NR-PDSCH Receiver:} At the receiver, the received signal undergoes timing synchronization and CP-OFDM demodulation to extract the NR-PDSCH transmitted information. Prior to NR-PDSCH decoding, phase noise estimation and correction based on MMSE estimation are applied to mitigate the impact of CPE on the received signal. The equations (\ref{Equ:03}) is  applied for that purpose.
\end{description}
\section{METHODOLOGY}
\label{sec:Methodology}

The methodology of this study involves evaluating the performance of 5G-NR mmWave communication systems, focusing on the assessment of phase noise models and the implementation of MMSE algorithms for CPE estimation and compensation. The simulation framework and setup parameters are presented to provide insights into the conducted experiments and assessments. The simulation framework leverages the advanced capabilities of the 5G Toolbox in MATLAB \cite{MathWorksR2022b}, offering a comprehensive environment for phase noise modeling and CPE estimation and correction in 5G-NR mmWave communication systems.

\subsection{Simulation Setup Parameters}
The experimental setup was executed using MATLAB \cite{MathWorksR2022b} which integrates the functionalities required for simulating the 5G-NR mmWave transmission chain. This tool enables the demonstration of phase noise impact in a 5G-NR mmWave system and the role of Phase Tracking Reference Signal (PT-RS) in compensating for Common Phase Error (CPE). Table \ref{Table:02}, Table \ref{Table:04}, Table \ref{Table:05} and Table \ref{Table:06} list the parameters used during the simulation process.

\begin{table}[htbp]
\centering
\caption{Global Simulation Parameters}
\resizebox{0.65\textwidth}{!}{
\begin{tabular}{|l|c|p{0.35\textwidth}|} \hline
\textbf{Global Simulation} & \centering \textbf{Values}  & \textbf{Description}  \\ \hline
NumFrames          & \centering 2  &  Simulation duration in number of frames \\ 
Fc  [GHz]          & \centering [29.55, 30, 60]   &  Carrier frequency \\ 
Nfft               & \centering 1024   &  IFFT/FFT size  \\ 
SampleRate [MHz]   & \centering 61.44   &  Sampling rate \\  \hline
\end{tabular}
}
\label{Table:02}
\end{table}

The carrier frequency for the simulation is set at 29.55 GHz, 30 GHz and 60 GHz, allowing for the exploration of different carrier frequencies' impact on the system's behavior, aligned with the mmWave spectrum characteristics.

\begin{table}[htbp]
\centering
\caption{Cell Parameters Configuration}
\resizebox{0.85\textwidth}{!}{
\begin{tabular}{|l|c|p{11cm}|} \hline
\textbf{Parameters}	        &\centering  \textbf{Values}	& \textbf{Description} \\ \hline
NCellID   	        &\centering 1 &	Network Cell Identifier, indicating the identity of the serving cell in a cellular network. \\ 
SubcarrierSpacing 	&\centering 60 [kHz]	&Subcarrier Spacing (SCS) [kHz]. \\ 
NSizeGrid	          &\centering  66	& The number of resource blocks (RB) in the grid. \\ 
SymbolsPerSlot	    &\centering   14	&  Indicates the number of symbols in each time slot.\\ 
\hline
\end{tabular}	
}	
\label{Table:04}
\end{table}

Table \ref{Table:04} defines the cell parameters used in the simulation for evaluating phase noise models and the Zero-Forcing/MMSE algorithms within the 5G-NR mmWave context.

\begin{table}[htbp]
\centering
\caption{NR-PDSCH Parameters Configuration}	
\label{Table:05}
\resizebox{0.85\textwidth}{!}{
\begin{tabular}{|l|c|p{0.7\textwidth}|} \hline
\textbf{Parameters}	&\centering \textbf{Values}	& \textbf{Description} \\ \hline
Modulation	&\centering 64QAM, 256QAM&	Indicates the modulation scheme being used.\\ 
NumLayers  	&\centering 1	&Specifies the number of layers. \\ 
MappingType	&\centering 'A'&	Refers to the mapping type.\\ 
PRBSetType	&\centering 'VRB'&	Indicates the Physical Resource Block (PRB) Set Type. VRB refers to Virtual Resource Block.\\ 
RNTI	      &\centering 2	& Radio Network Temporary Identifier (RNTI). \\ 
EnablePTRS	&\centering 1 &	PTRS (Phase-Tracking Reference Signal) is enabled (set to 1).\\ 
\hline
\end{tabular}		
}
\end{table}

Table \ref{Table:05} provides the configuration parameters specific to the Physical Downlink Shared Channel (PDSCH) within the simulation setup, with PTRS enabled to ensure accurate phase noise tracking.

\begin{table}[htbp]
\centering
\caption{DM-RS and PT-RS Parameters Configuration}
\label{Table:06}
\resizebox{0.85\textwidth}{!}{
\begin{tabular}{|l|c|p{0.65\textwidth}|} \hline
\textbf{Parameters}	            &\centering  \textbf{Values}	& \textbf{Description} \\ \hline
DMRS-ConfigurationType	&\centering 1&	Specifies the type of DM-RS configuration.\\
DMRS-ReferencePoint	    &\centering 'CRB0'&	Defines the reference point for DM-RS generation.\\
DMRS-TypeAPosition	    &\centering 2&	Indicates the position of DM-RS type A in the resource grid.\\
DMRS-AdditionalPosition	&\centering 0&	Specifies any additional position of DM-RS.\\
DMRS-Length	            &\centering 1&	Indicates the length of DM-RS.\\
PTRS-TimeDensity	      &\centering 1&	This parameter sets the time density of the PT-RS.\\
PTRS-FrequencyDensity	  &\centering 2&	This parameter sets the frequency density of the PT-RS.\\
PTRS-REOffset	          &\centering '00'&	This parameter sets the Resource Element (RE) offset for the PT-RS.\\
\hline
\end{tabular}
}		
\end{table}

\subsection{Phase Noise Model Evaluation}

The evaluation of phase noise models 'A' \cite{Samsung2016}, 'B' \cite{Samsung2016} and 'C' (TR 38.803 \cite{38.8032022}) is based on the parameter sets defined in Table \ref{Table:01}. These models are implemented within MATLAB's 5G Toolbox \cite{MathWorksR2022b} using the \verb|hPhaseNoiseModel()| function which simulates the phase noise characteristic PSD for specific frequency offsets and carrier frequencies.

Error Vector Magnitude (EVM) in both dB and percentage is the key metric for comparison, used to assess phase error variance introduced by each model. The models' accuracy is tested across diverse channel conditions, including scenarios with high path loss, multipath fading and varying SNR, ensuring comprehensive evaluation under realistic mmWave conditions.

\subsection{MMSE Algorithms Implementation}
In this subsection, we detail the implementation of the  MMSE algorithms for CPE estimation and compensation in the context of 5G-NR mmWave. These algorithms are crucial for mitigating phase errors introduced by various factors, thereby enhancing the reliability and performance of mmWave communication systems.

The MMSE algorithms for CPE estimation and compensation are derived from equations (\ref{Equ:03}) and (\ref{Equ:04}). The implementation of these algorithms within the NR-PDSCH receiver is executed using MATLAB's \verb|nrEqualizeMMSE()| function, designed to handle the complexities of CPE estimation and compensation in 5G-NR mmWave systems.

Performance metrics such as Block Error Rate (BLER), Signal-to-Noise Ratio (SNR) and throughput are employed to assess the effectiveness of the MMSE algorithms. These metrics serve as key indicators of the robustness and reliability of the algorithms for CPE estimation and compensation in the 5G-NR mmWave context.

\subsection{NR-PDSCH Receiver Implementation}
The NR-PDSCH receiver implementation utilizes the capabilities of MATLAB's 5G Toolbox \cite{MathWorksR2022b}, leveraging a suite of essential functions tailored for 5G communication systems. A selection of these functions, along with their descriptions, is provided in Table \ref{Table:03}.

\begin{table}[htbp]
\centering
\caption{5G Toolbox Functions Employed in NR-PDSCH Receiver}
\resizebox{0.85\textwidth}{!}{
\begin{tabular}{|l|p{0.7\textwidth}|} \hline
%\textbf{5G Toolbox Functions} 	& \textbf{Description} \\  \hline
\textbf{5G Toolbox Functions} 	& \textbf{Description} \\  \hline
nrChannelEstimate()	& Performs channel estimation, returning channel estimate H and noise variance estimate. \\  
nrEqualizeMMSE()      & Executes MMSE equalization on extracted resource elements of NR-PDSCH using estimated channel and noise variance. \\  
demodulate()    &  Executes Quadrature Amplitude Modulation (M-QAM) demodulation of the received signal. \\  
nrTimingEstimate()   & Performs timing estimation by cross-correlating known reference signals with the received signal. \\  
nrSymbolDemodulate()   & Demodulates symbols into bits according to the constellation. \\  
\hline
\end{tabular}	
}	
\label{Table:03}
\end{table}

By leveraging these functions within the 5G Toolbox, the NR-PDSCH receiver implementation achieves accurate and efficient processing of received signals by facilitating the recovery of transmitted data in 5G-NR mmWave communication systems.

\section{RESULTS AND DISCUSSION}
\label{sec:ResultsAndDiscussion}

This section presents and analyzes the results obtained from the evaluation of phase noise models and the implementation of CPE compensation using MMSE algorithms in 5G-NR mmWave communication systems. The discussion focuses on the impact of these techniques across various modulation schemes, SNR levels, antenna configurations and phase noise models by highlighting the improvements in system performance and the practical implications for real-world communication scenarios.

\subsection{CPE Compensation Performance for Various Modulation Schemes}

\begin{figure}[htbp]
	\centering
	\fbox{
\includegraphics[page = 2,clip, trim=0.0cm 0.0cm 0.0cm 0.0cm, width=0.85\textwidth]{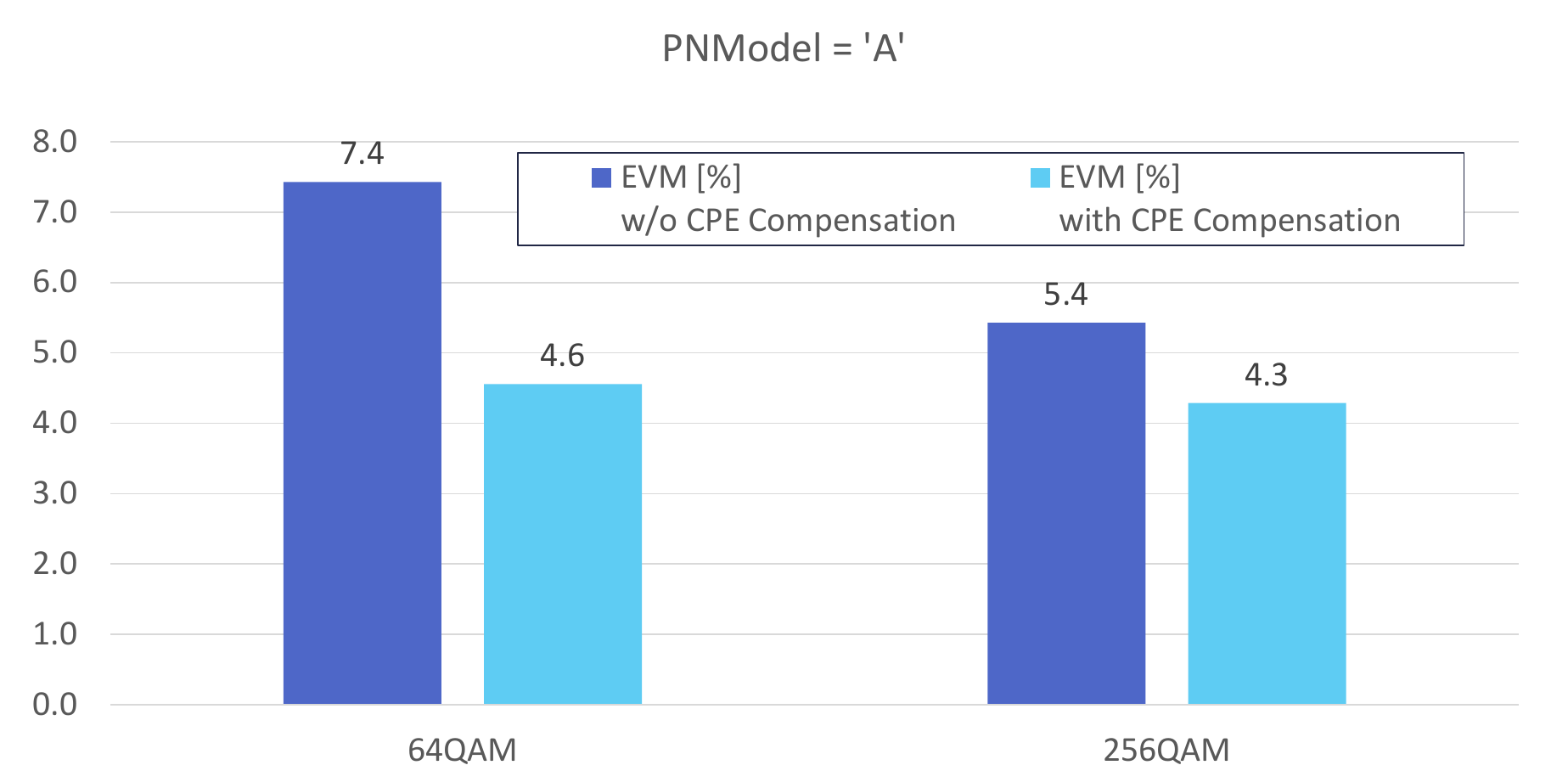}
}
	\caption{Impact of CPE Compensation on EVM [\%] for 64QAM and 256QAM under PN Model 'A'}
	\label{Fig:04}
\end{figure}

\begin{figure}[htbp]
	\centering
	\fbox{
\includegraphics[page = 3,clip, trim=0.0cm 0.0cm 0.0cm 0.0cm, width=0.85\textwidth]{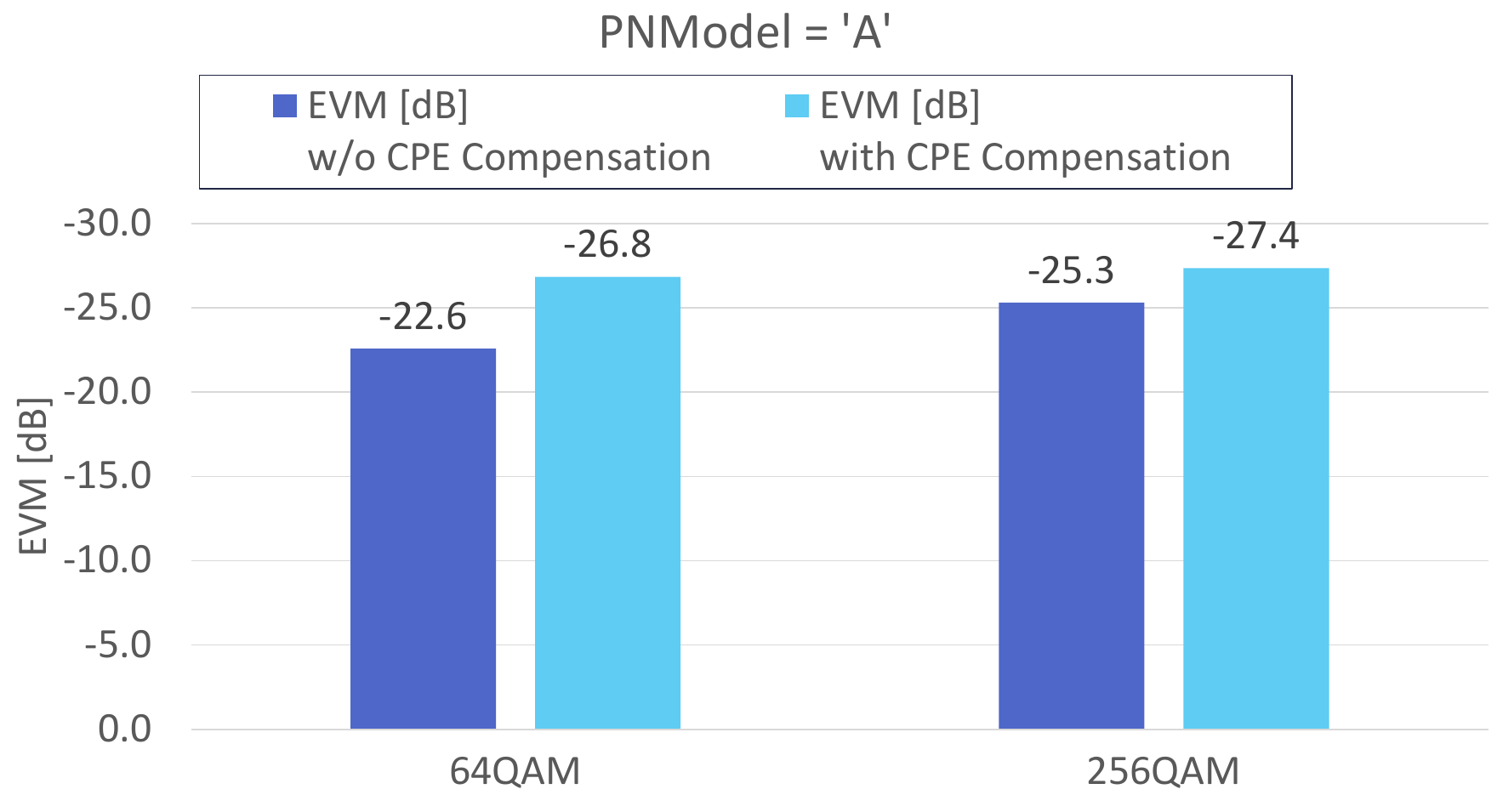}
}
	\caption{Impact of CPE Compensation on EVM [dB] for 64QAM and 256QAM under PN Model 'A'}
	\label{Fig:05}
\end{figure}

\begin{figure}[htbp]
	\centering
	\fbox{
\includegraphics[page = 4,clip, trim=0.0cm 0.0cm 0.0cm 0.0cm, width=0.85\textwidth]{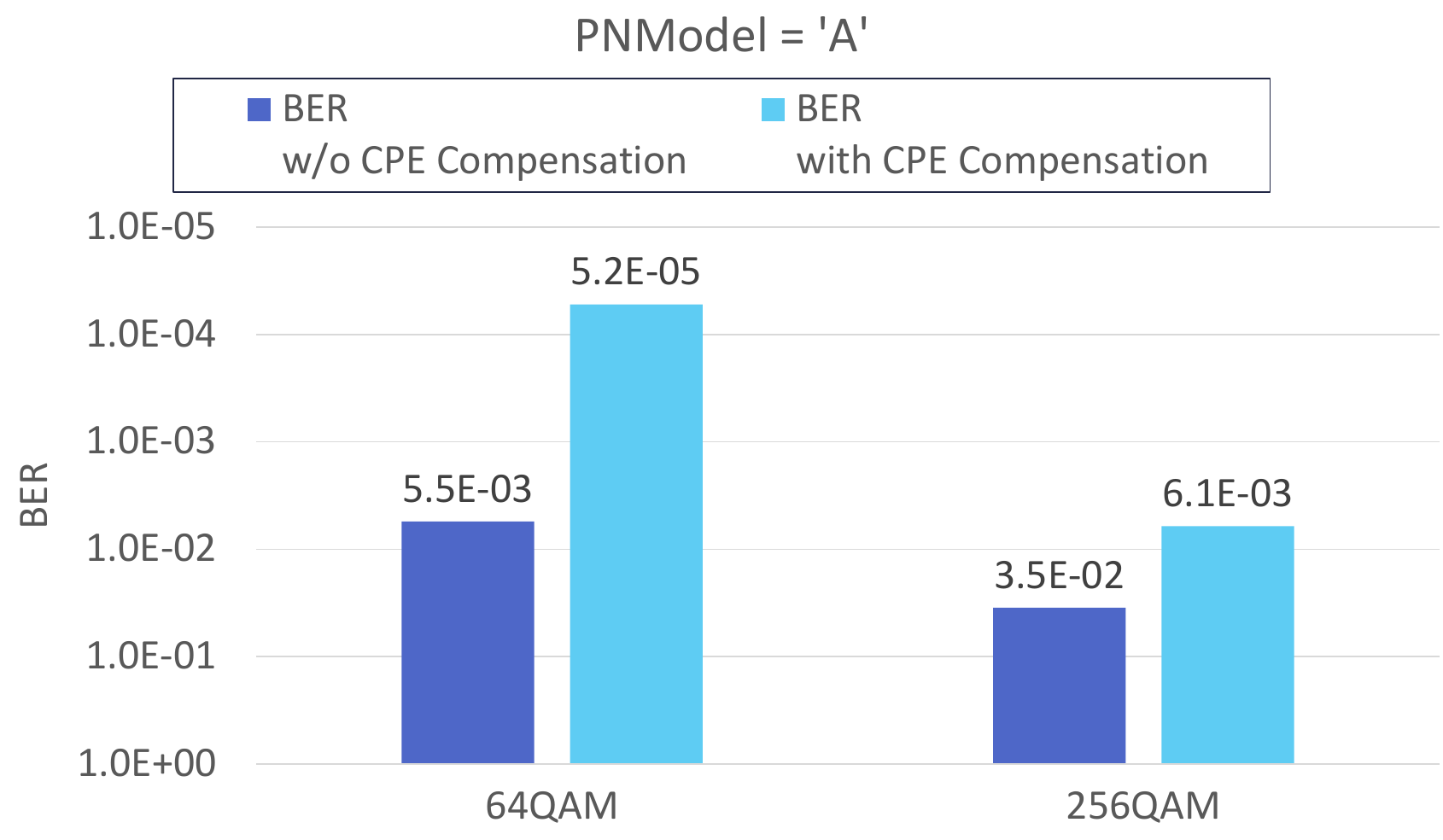}
}
	\caption{Impact of CPE Compensation on BER for 64QAM and 256QAM under PN Model 'A'}
	\label{Fig:06}
\end{figure}

Figures \ref{Fig:04}, \ref{Fig:05} and \ref{Fig:06} demonstrate the impact of Common Phase Error (CPE) compensation on the performance of 64QAM and 256QAM modulation schemes. For 64QAM, CPE compensation reduces the Error Vector Magnitude (EVM) from 7.4\% to 4.6\% and improves the EVM in dB from -22.6 dB to -26.8 dB, indicating a significant enhancement in signal quality. The Bit Error Rate (BER) for 64QAM also drops dramatically from \(5.5 \times 10^{-3}\) to \(5.2 \times 10^{-5}\), highlighting a substantial reduction in transmission errors.

Similarly, for 256QAM, CPE compensation decreases the EVM from 5.4\% to 4.3\% and improves the EVM in dB from -25.3 dB to -27.4 dB. The BER for 256QAM also shows improvement, although less pronounced, decreasing from \(3.5 \times 10^{-2}\) to \(6.1 \times 10^{-3}\). These results demonstrate that CPE compensation significantly enhances signal integrity and reduces errors, making it a valuable technique in communication systems utilizing these modulation schemes.

\subsection{CPE Compensation Performance Across Various SNR Levels, Phase Noise Models and Antenna Configurations}

\begin{figure}[htbp]
	\centering
	\fbox{
\includegraphics[page = 5,clip, trim=0.0cm 0.0cm 0.0cm 0.0cm, width=0.85\textwidth]{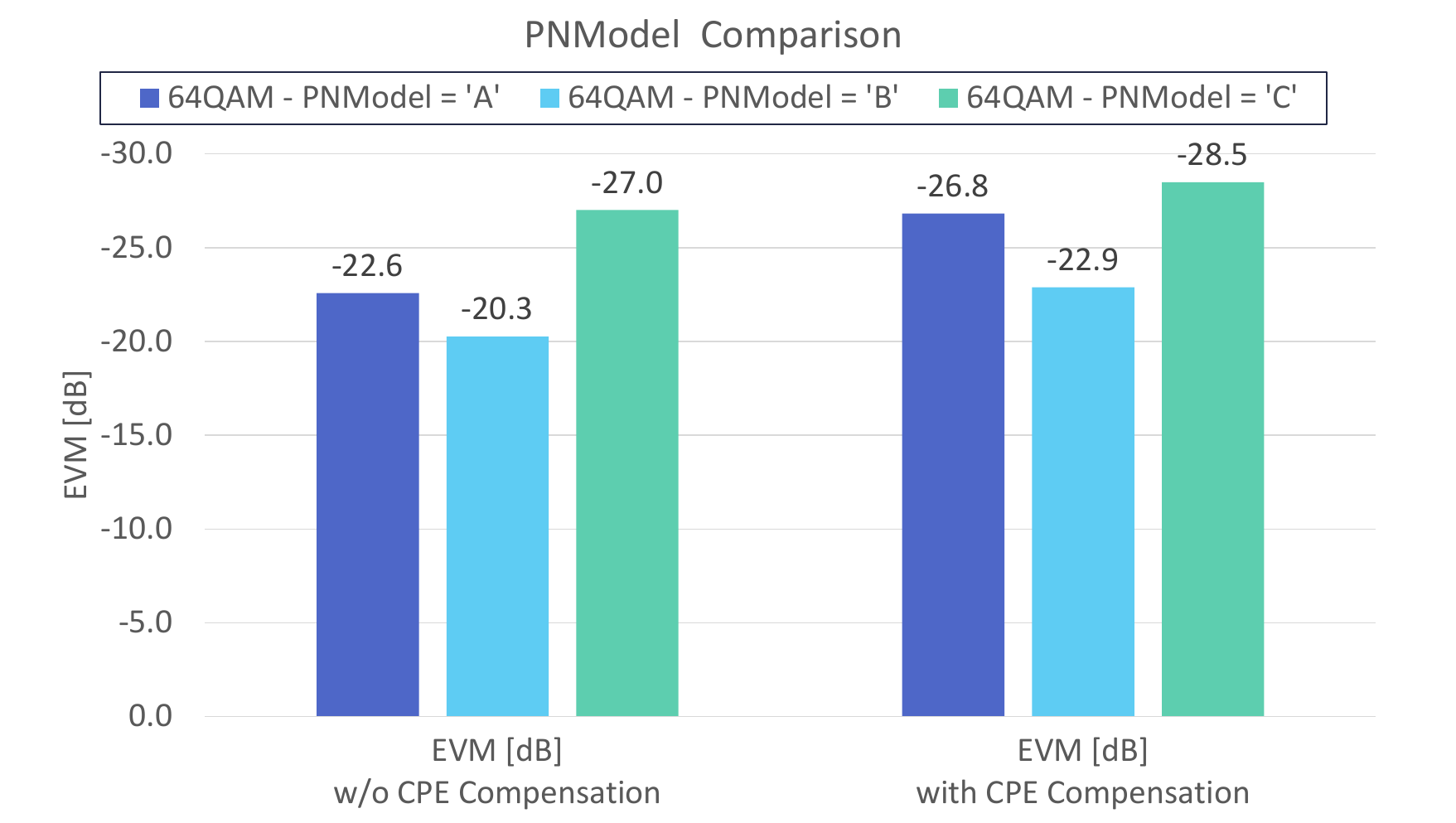}
}
	\caption{Comparison of EVM [dB] for 64QAM Across Different PN Models with and without CPE Compensation}
	\label{Fig:07}
\end{figure}

\begin{figure}[htbp]
	\centering
	\fbox{
\includegraphics[page = 6,clip, trim=0.0cm 0.0cm 0.0cm 0.0cm, width=0.85\textwidth]{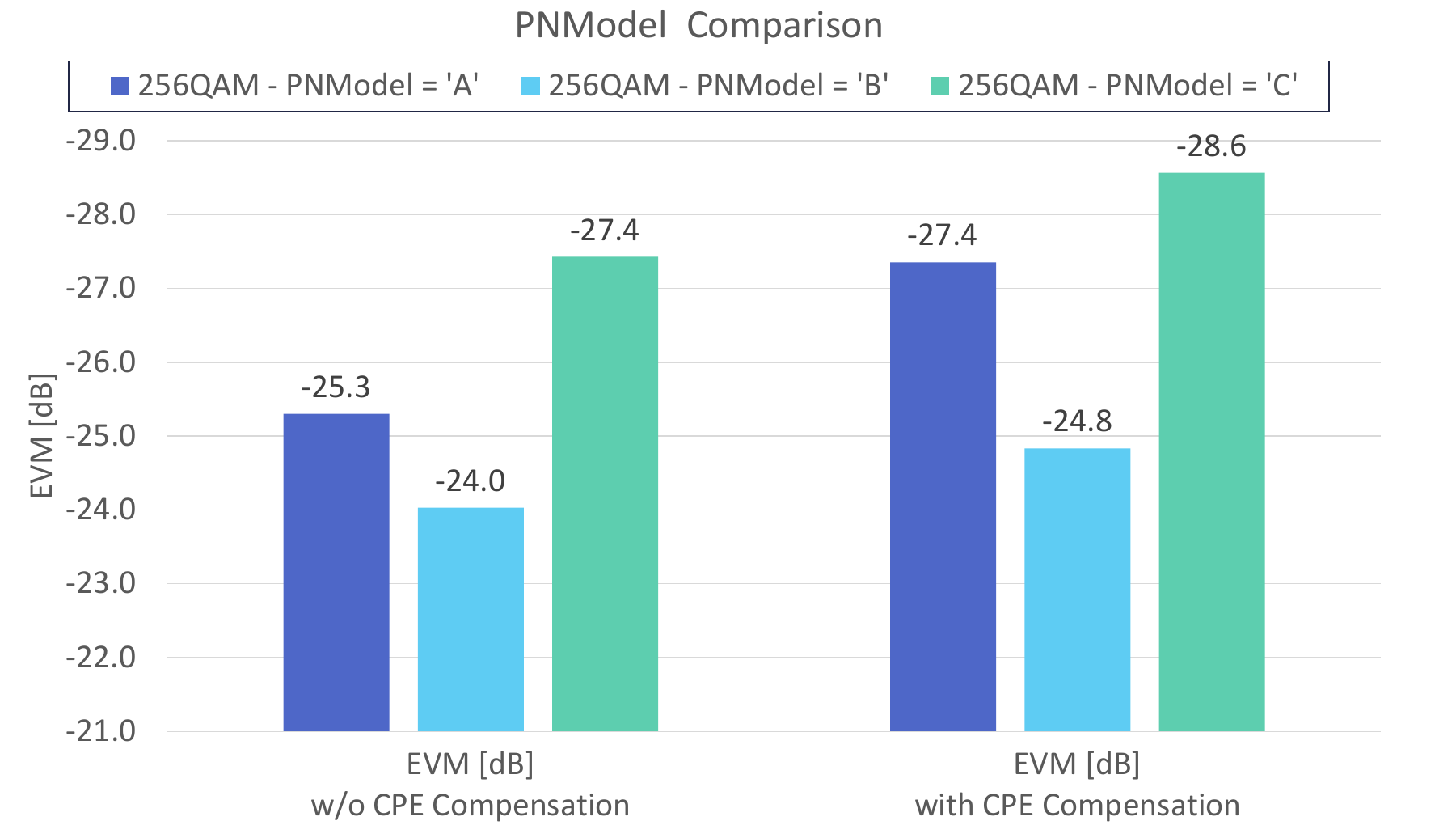}
}
	\caption{Comparison of EVM [dB] for 256QAM Across Different PN Models with and without CPE Compensation}
	\label{Fig:08}
\end{figure}

This section analyzes CPE Compensation performance across different SNR levels, Phase Noise models and antenna configurations. Figures \ref{Fig:07} and \ref{Fig:08} show the Error Vector Magnitude (EVM) values in decibels (dB) for 64QAM and 256QAM modulation schemes under various Phase Noise (PN) models, both with and without CPE compensation. EVM measures the accuracy of the modulation and demodulation processes, with lower values indicating better performance. The results consistently show that CPE compensation improves EVM across all modulation schemes and PN models. For instance, in the case of 64QAM with PN Model 'A', the EVM improves from -22.6 dB without compensation to -26.8 dB with compensation. This trend is consistent across most scenarios, indicating the effectiveness of CPE compensation techniques in mitigating phase errors and enhancing communication performance in mmWave systems.

Variations in performance among different PN models and modulation schemes are also observed. PN Model 'B' generally exhibits lower EVM values compared to PN Model 'C', suggesting that the choice of PN model significantly impacts system performance. 64QAM and 256QAM also exhibit differences in EVM, with 256QAM typically showing higher EVM values due to its increased complexity and higher data rate.

When analyzing the performance of Phase Noise (PN) Models 'A', 'B' and 'C' under the same modulation scheme, PN Model 'B' generally performs better than PN Models 'A' and 'C' in terms of EVM values, both with and without CPE compensation. This indicates that PN Model 'B' is more effective at reducing phase errors and maintaining signal accuracy, leading to improved communication performance.

\subsection{BLER Performance Under Different Antenna Configurations}

\begin{figure}[htbp]
	\centering
	\fbox{
\includegraphics[page = 7,clip, trim=0.0cm 0.0cm 0.0cm 0.0cm, width=0.85\textwidth]{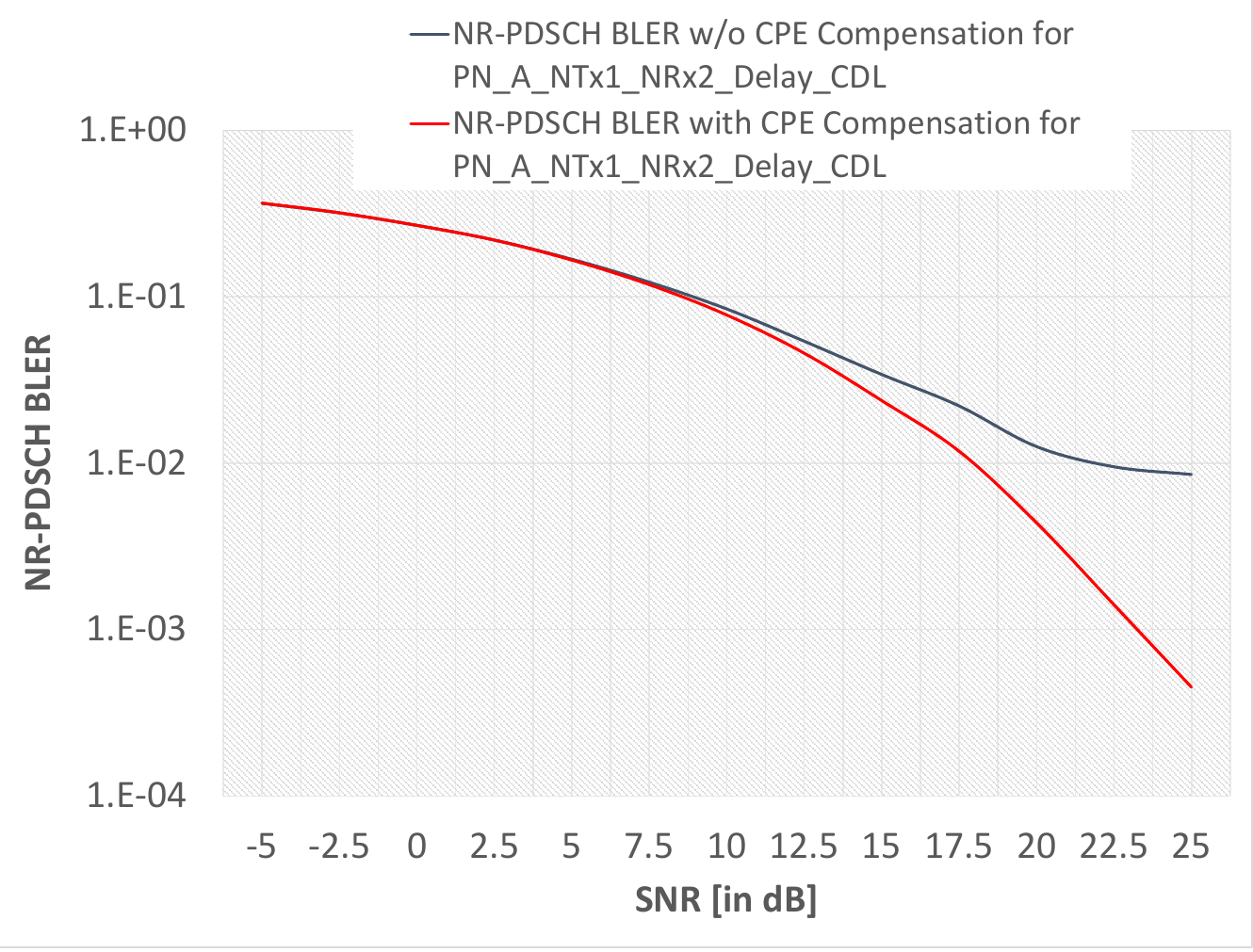}
}
	\caption{NR-PDSCH BLER Performance Without CPE Compensation for PN\_A\_NTx1\_NRx2\_Delay\_CDL}
	\label{Fig:09}
\end{figure}

\begin{figure}[htbp]
	\centering
	\fbox{
\includegraphics[page = 8,clip, trim=0.0cm 0.0cm 0.0cm 0.0cm, width=0.85\textwidth]{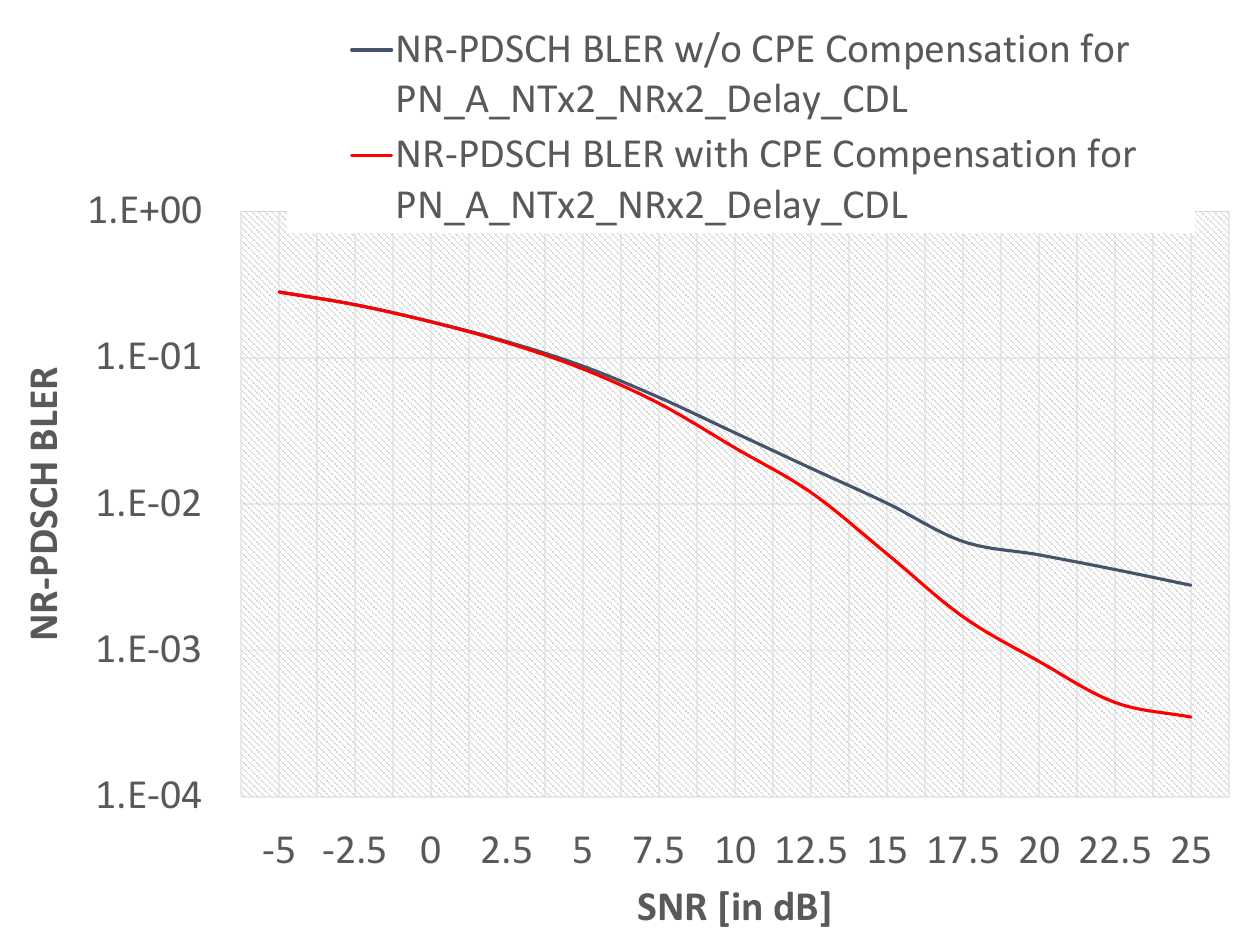}
}
	\caption{NR-PDSCH BLER Performance Without CPE Compensation for PN\_A\_NTx2\_NRx2\_Delay\_CDL}
	\label{Fig:10}
\end{figure}

Figures \ref{Fig:09} and \ref{Fig:10} illustrate the BLER performance of NR-PDSCH under different antenna configurations (NTx = 1 and 2, NRx = 2) and phase noise models. The results consistently show that CPE compensation enhances system performance by significantly lowering the BLER across various SNR values, particularly at higher SNR levels where the gap between compensated and non-compensated curves widens. This indicates that CPE compensation is highly effective in mitigating the adverse effects of phase noise, leading to more reliable data transmission.

\begin{figure}[htbp]
	\centering
	\fbox{
\includegraphics[page = 9,clip, trim=0.0cm 0.0cm 0.0cm 0.0cm, width=0.85\textwidth]{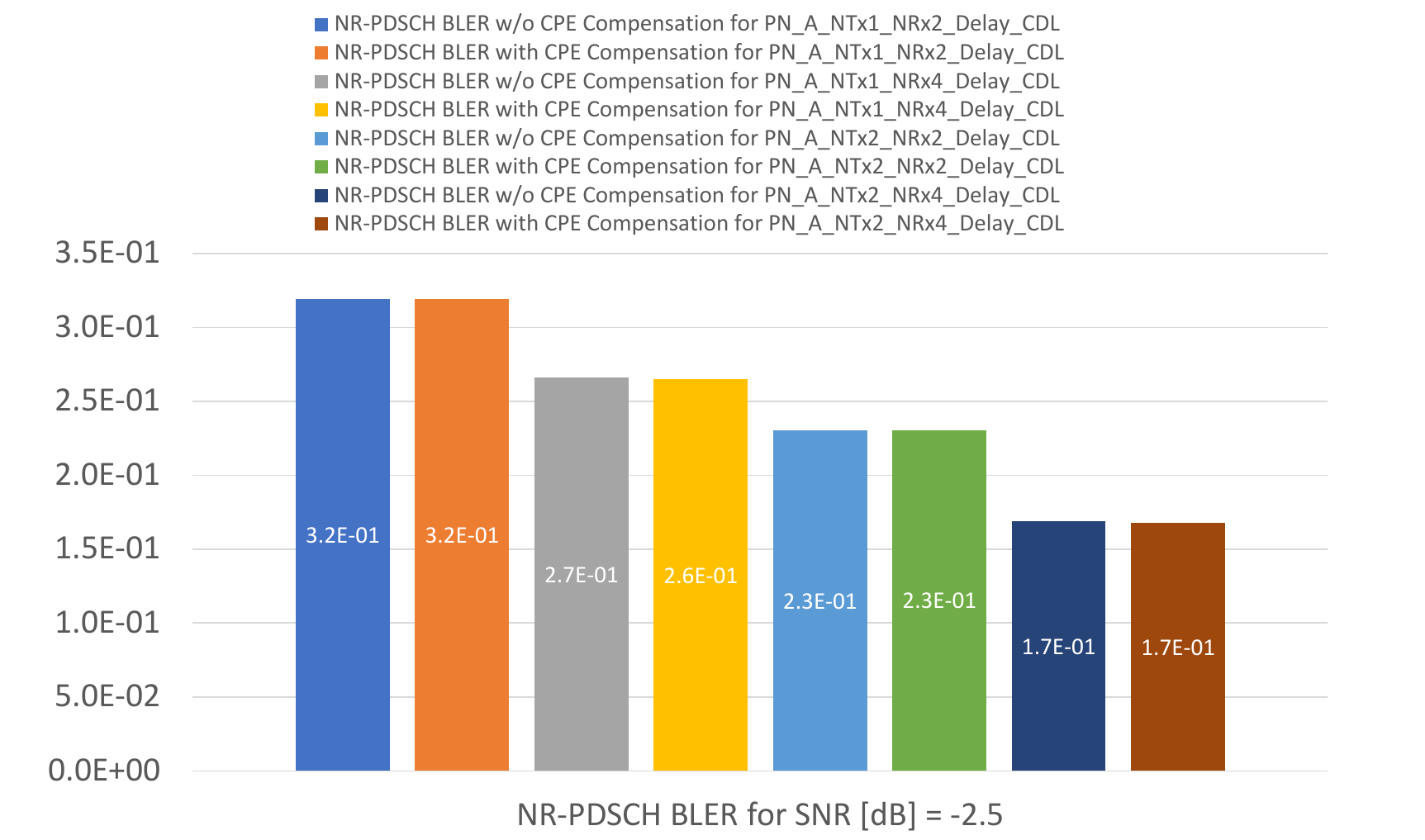}
}
	\caption{NR-PDSCH BLER for Various Antenna Configurations With and Without CPE Compensation at SNR = -2.5 dB}
	\label{Fig:11}
\end{figure}

\begin{figure}[htbp]
	\centering
	\fbox{
\includegraphics[page = 10,clip, trim=0.0cm 0.0cm 0.0cm 0.0cm, width=0.85\textwidth]{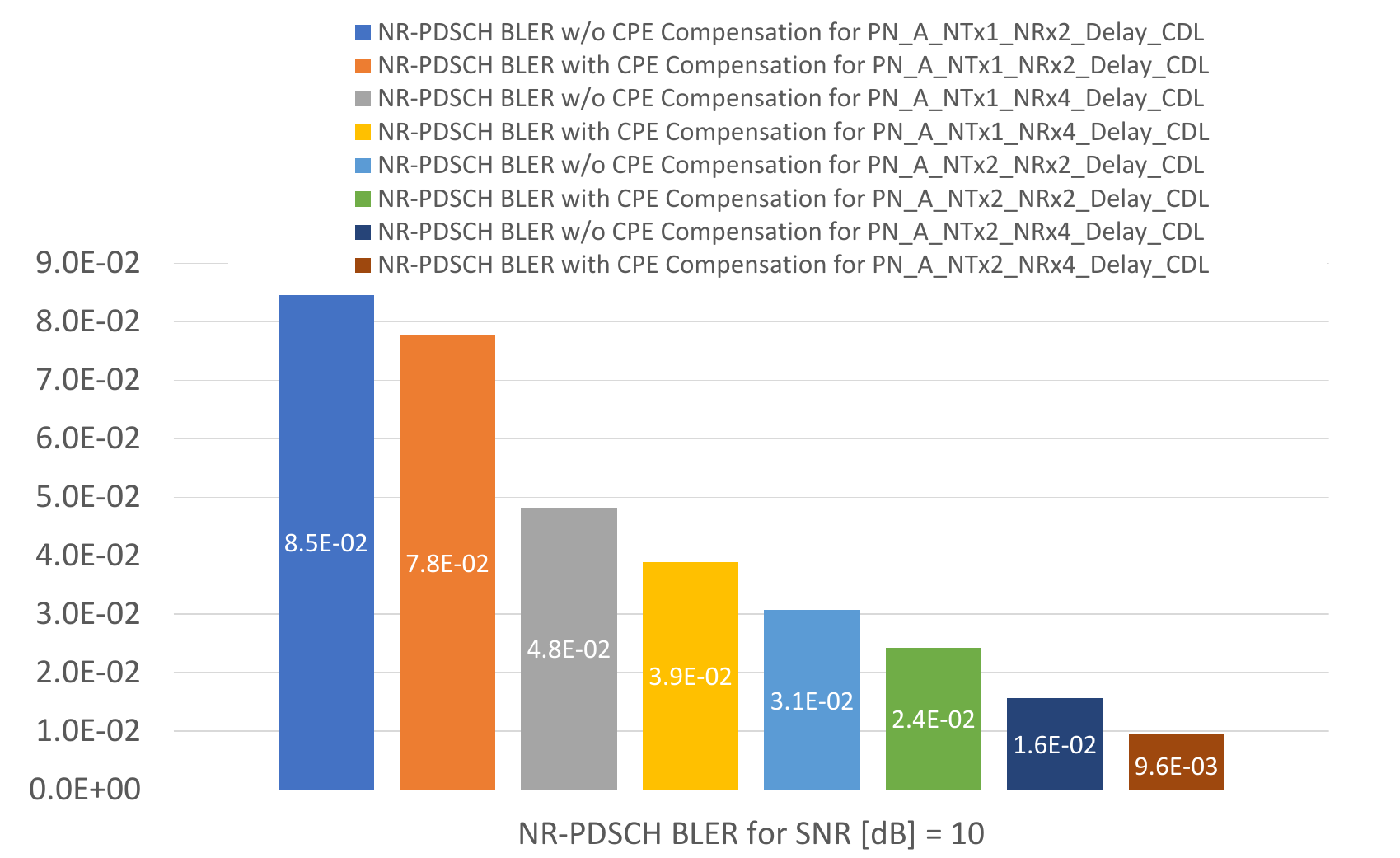}
}
	\caption{NR-PDSCH BLER for Various Antenna Configurations With and Without CPE Compensation at SNR = 10 dB}
	\label{Fig:12}
\end{figure}

Figures \ref{Fig:11} and \ref{Fig:12} present BLER performance for NR-PDSCH at SNR values of -2.5 dB and 10 dB, respectively, under different antenna configurations. At lower SNR (-2.5 dB), BLER remains relatively high across all configurations, with marginal differences observed between cases with and without CPE compensation. At this low SNR, the impact of CPE compensation on BLER is limited due to the dominant noise floor. However, at higher SNR (10 dB), a pronounced reduction in BLER with CPE compensation is observed across all configurations. The most significant improvements are noted in scenarios with more antennas (NRx = 4), indicating that CPE compensation becomes increasingly effective as the number of antennas increases, due to improved channel estimation and spatial diversity.

\subsection{Comparison of PN Models With and Without CPE Compensation}

\begin{figure}[htbp]
	\centering
	\fbox{
\includegraphics[page = 11,clip, trim=0.0cm 0.0cm 0.0cm 0.0cm, width=0.85\textwidth]{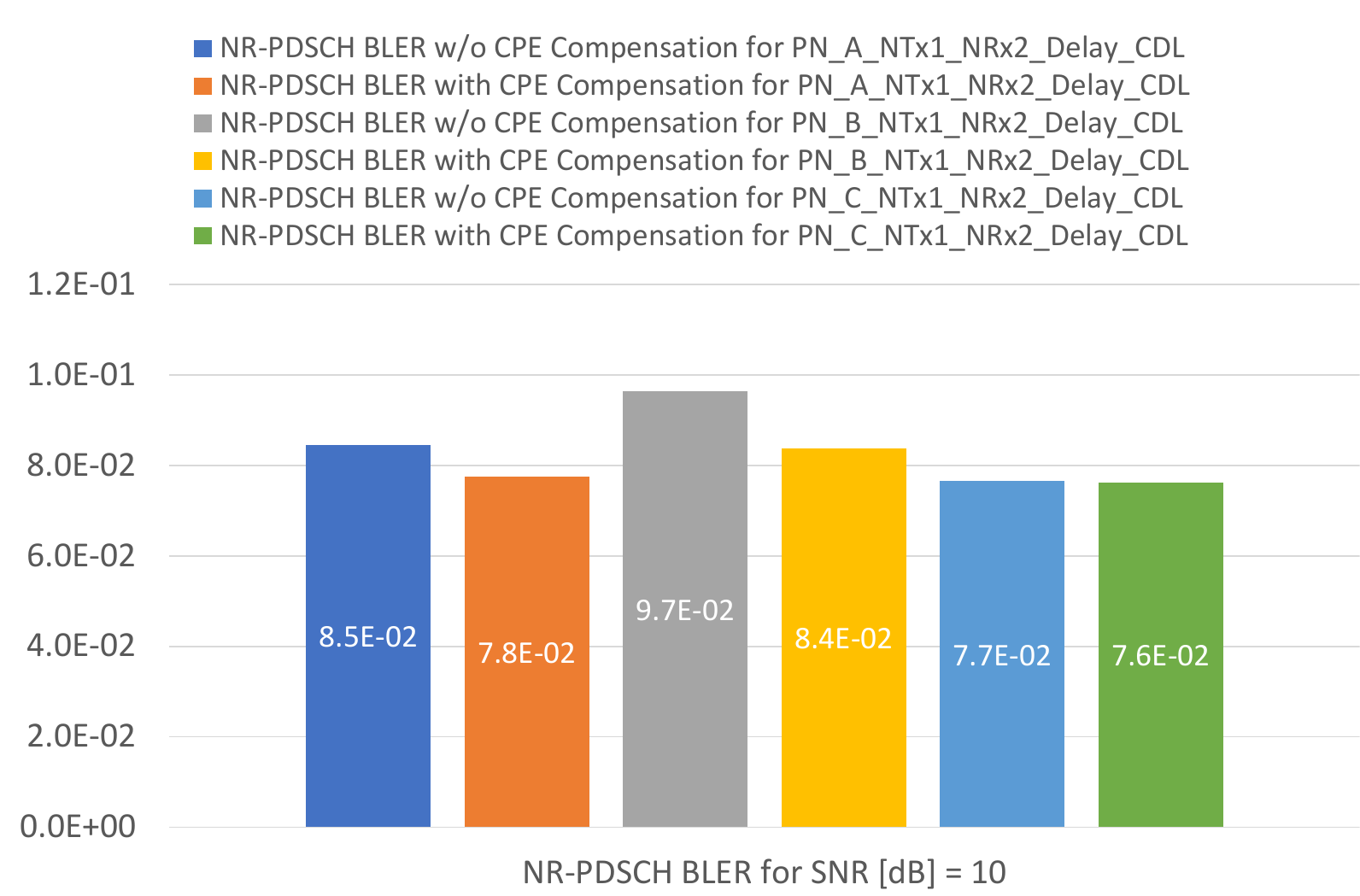}
}
	\caption{NR-PDSCH BLER for PN Models 'A', 'B' and 'C', With and Without CPE Compensation at SNR = 10 dB}
	\label{Fig:13}
\end{figure}

The BLER results depicted in Figure \ref{Fig:13} indicate that CPE compensation consistently improves performance across all phase noise models. The reductions in BLER are as follows: from $8.5\times 10^{-2}$ to $7.8\times 10^{-2}$ for PN Model 'A', from $9.7\times 10^{-2}$ to $8.4\times 10^{-2}$ for PN Model 'B' and from $8.4\times 10^{-2}$ to $7.6\times 10^{-2}$ for PN Model 'C'. Among these, PN Model 'B' shows the highest BLER, both with and without CPE compensation, indicating that it introduces more severe phase noise compared to PN Models 'A' and 'C'. Conversely, PN Model 'C', when CPE compensation is applied, exhibits the lowest BLER, showcasing the most significant improvement.

The variations in BLER performance among the PN models highlight the differing severity and predictability of the phase noise inherent in each model. PN Model 'B' presents a greater challenge for compensation algorithms, leading to higher BLER. On the other hand, PN Model 'C', with its less severe and more predictable phase noise, allows compensation algorithms to perform more effectively, resulting in the most significant improvement and the lowest BLER.

\subsection{Discussion of Results and Practical Implications}

The results of this study underscore the critical role of CPE compensation in enhancing the performance of 5G-NR mmWave communication systems. Across various modulation schemes, antenna configurations and phase noise models, the implementation of CPE compensation consistently leads to improvements in signal quality, as evidenced by reductions in EVM and BLER. The effectiveness of CPE compensation becomes particularly evident at higher SNR levels, where it significantly reduces error rates, thereby enhancing the reliability and robustness of communication systems.

Moreover, the comparison of different PN models reveals that the severity and predictability of phase noise have a substantial impact on the performance of CPE compensation techniques. Models with more predictable phase noise, such as PN Model 'C', allow for more effective compensation, resulting in lower BLER and improved system performance.

The practical implications of these findings are significant for the design and optimization of 5G-NR mmWave communication systems. The choice of phase noise model and the implementation of effective CPE compensation strategies are critical for achieving optimal system performance, particularly in environments characterized by high SNR and complex antenna configurations. Future research should continue to explore the characteristics and limitations of various phase noise models and compensation techniques to further enhance the reliability and efficiency of mmWave communication systems.

\section{CONCLUSION}
\label{sec:Conclusion}

In this study, we evaluated the performance of Common Phase Error (CPE) estimation and compensation techniques in 5G-NR mmWave communication systems across various phase noise models and modulation schemes. The key findings indicate that CPE compensation significantly improves signal quality, with EVM reductions from 7.4\% to 4.6\% for 64QAM and from 5.4\% to 4.3\% for 256QAM, alongside notable reductions in BER. These improvements demonstrate the effectiveness of CPE compensation in enhancing communication reliability, particularly under challenging noise environments.

The practical implications of these findings are substantial for the design and optimization of 5G-NR mmWave systems. CPE compensation proves to be crucial in scenarios involving higher SNR values and complex antenna configurations, where the benefits of enhanced spatial diversity and improved channel estimation are most evident. This underscores the importance of incorporating robust CPE compensation techniques in future 5G deployments to ensure optimal system performance.

Looking ahead, future research should focus on refining phase noise models and exploring advanced compensation techniques to further enhance system resilience. Investigating the interplay between different PN models and the effectiveness of compensation algorithms in various real-world scenarios will be key to developing more sophisticated and reliable 5G-NR mmWave communication systems.

Overall, this study highlights the vital role of CPE compensation in mitigating phase errors, thereby contributing to the advancement of next-generation wireless networks.

\section{THE PERSPECTIVES}

The ongoing advancements in 5G-NR and mmWave communication technologies continue to push the boundaries of wireless communication performance. This study demonstrates that the implementation of CPE compensation significantly enhances signal quality by reducing EVM and BLER across various modulation schemes, phase noise models and antenna configurations.

Looking forward, several key perspectives can be drawn:
\begin{itemize}
	\item (i) Future research should focus on developing more advanced CPE compensation algorithms that adapt to diverse phase noise environments. The integration of machine learning and artificial intelligence techniques holds promise for optimizing these algorithms for real-time applications \cite{Kim2021}.
	\item (ii) The deployment of massive MIMO and beamforming techniques in 5G-NR systems presents new opportunities for improving CPE compensation. These technologies can enhance spatial diversity and provide more robust phase noise mitigation.
	\item (iii) As communication systems evolve towards higher frequency bands, such as sub-THz and THz, the challenges of phase noise will become more pronounced. Research should investigate the scalability of CPE compensation techniques in these higher frequency domains \cite{Zhang2022}.
	\item (iv) Ensuring interoperability between different vendors' equipment and adherence to standardized protocols is crucial for the widespread adoption of CPE compensation technologies. Collaborative efforts between industry and academia can facilitate the development of universally accepted standards.
	\item (v) Extensive field trials and real-world deployments are essential to validate the theoretical and simulation-based findings. Practical insights gained from these deployments can inform further refinements of CPE compensation strategies.
\end{itemize}

\nocite{*}
%\printbibliography[]
\printbibliography[heading=bibintoc,title={\textsc{REFERENCES}},env=bibliography]

\end{document}